\documentclass[conference,hidelinks]{IEEEtran}

\usepackage{graphicx,url}
\usepackage[english]{babel}
\addto\captionsenglish{}
\usepackage[utf8]{inputenc}
\usepackage{amsmath}
\usepackage{amssymb}
\usepackage{lipsum}
\usepackage{mathtools}
\usepackage{cuted}
\usepackage{cite}
\usepackage{graphicx}
\usepackage{tikz}
\usepackage{pgfplots}
\usepackage{tikzscale} 
\pgfplotsset{compat=newest} 
\usepackage{scalefnt}
 
\usepackage{orcidlink}
\usepackage{dblfloatfix}
\usepackage[acronym,shortcuts]{glossaries}
\usepackage{algorithm, algpseudocode}
\usepackage{algcompatible}
\usepackage{bm}
\usepackage{makecell}
\usepackage{hyperref}
\usepackage{xcolor}
\usepackage{tcolorbox}
\usepackage{multirow}
\usepackage{subcaption}
\usepackage{textcomp}
\usepackage[normalem]{ulem}

\newacronym{RMSE}{RMSE}{root mean square error}
\newacronym{i.i.d.}{i.i.d.}{independent, identically distributed}
\newacronym{PA}{PA}{power amplifier}
\newacronym{SSPA}{SSPA}{solid-state power amplifier}
\newacronym{IBO}{IBO}{input back-off}
\newacronym{PSLR}{PSLR}{peak sidelobe level ratio}
\newacronym{MMSE}{MMSE}{minimum mean square error}
\newacronym{MF}{MF}{matched filter}
\newacronym{RPE}{RPE}{radar parameter estimation}
\newacronym{OTFS}{OTFS}{orthogonal time frequency space}
\newacronym{AFDM}{AFDM}{affine frequency division multiplexing}
\newacronym{MIMO}{MIMO}{multiple-input multiple-output}
\newacronym{SISO}{SISO}{single-input single-output}
\newacronym{ISAC}{ISAC}{integrated sensing and communications}
\newacronym{3D}{3D}{three-dimensional}
\newacronym{2D}{2D}{two-dimensional}
\newacronym{1D}{1D}{one-dimensional}
\newacronym{RX}{RX}{receiver}
\newacronym{TX}{TX}{transmitter}
\newacronym{BF}{BF}{beamforming}
\newacronym{mmWave}{mmWave}{millimeter-wave}
\newacronym{SotA}{SotA}{state-of-the-art}
\newacronym{ULA}{ULA}{uniform linear array}
\newacronym{QAM}{QAM}{quadrature amplitude modulation}
\newacronym{ISFFT}{ISFFT}{inverse symplectic finite Fourier transform}
\newacronym{SFFT}{SFFT}{symplectic finite Fourier transform}
\newacronym{AWGN}{AWGN}{additive white Gaussian noise}
\newacronym{OFDM}{OFDM}{orthogonal frequency division multiplexing}
\newacronym{OCDM}{OCDM}{orthogonal chirp division multiplexing}
\newacronym{BS}{BS}{base station}
\newacronym{UE}{UE}{user equipment}
\newacronym{DFT}{DFT}{discrete Fourier transform}
\newacronym{IDFT}{IDFT}{inverse discrete Fourier transform}
\newacronym{IFFT}{IFFT}{inverse fast Fourier transform}
\newacronym{TD}{TD}{time-domain}
\newacronym{wlg}{wlg}{without loss of generality}
\newacronym{CP}{CP}{cyclic prefix}
\newacronym{DAFT}{DAFT}{discrete affine Fourier transform}
\newacronym{DAF}{DAF}{discrete affine Fourier}
\newacronym{IDAFT}{IDAFT}{inverse discrete affine Fourier transform}
\newacronym{CPP}{CPP}{\textit{chirp-periodic} prefix}
\newacronym{IDZT}{IDZT}{inverse discrete Zak transform}
\newacronym{DZT}{DZT}{discrete Zak transform}
\newacronym{ICI}{ICI}{inter-carrier interference}
\newacronym{BER}{BER}{bit error rate}
\newacronym{DoF}{DoF}{degrees-of-freedom}
\newacronym{FD}{FD}{full-duplex}
\newacronym{SIMO}{SIMO}{single-input multiple-output}
\newacronym{MISO}{MISO}{multiple-input single-output}
\newacronym{AoD}{AoD}{angle-of-departure}
\newacronym{AoA}{AoA}{angle-of-arrival}
\newacronym{RF}{RF}{radio frequency}
\newacronym{SIM}{SIM}{stacked intelligent metasurfaces}
\newacronym{FPGA}{FPGA}{field programmable gate array}
\newacronym{UPA}{UPA}{uniform planar array}
\newacronym{CC}{CC}{communication-centric}
\newacronym{I/O}{I/O}{input-output}
\newacronym{iid}{i.i.d.}{independent and identically distributed}
\newacronym{IoT}{IoT}{internet of things}
\newacronym{V2X}{V2X}{vehicle-to-everything}
\newacronym{NTN}{NTN}{non-terrestrial network}
\newacronym{LEO}{LEO}{low-earth orbit}
\newacronym{THz}{THz}{terahertz}
\newacronym{EM}{EM}{expectation maximization}
\newacronym{RIS}{RIS}{reconfigurable intelligent surface}
\newacronym{DoA}{DoA}{direction-of-arrival}
\newacronym{DD}{DD}{doubly-dispersive}
\newacronym{ODDM}{ODDM}{orthogonal delay-Doppler division multiplexing}
\newacronym{LoS}{LoS}{line-of-sight}
\newacronym{NLoS}{NLoS}{non-line-of-sight}
\newacronym{6G}{6G}{sixth generation}
\newacronym{MPDD}{MPDD}{metasurfaces-parametrized DD}
\newacronym{GaBP}{GaBP}{Gaussian belief propagation}
\newacronym{MSE}{MSE}{mean-squared-error}
\newacronym{sIC}{soft IC}{soft interference cancellation}
\newacronym{soft RG}{soft RG}{soft replica generation}
\newacronym{BG}{BG}{belief generation}
\newacronym{SGA}{SGA}{scalar Gaussian approximation}
\newacronym{CLT}{CLT}{central limit theorem}
\newacronym{PDF}{PDF}{probability density function}
\newacronym{QPSK}{QPSK}{quadrature phase-shift keying}
\newacronym{OQAM}{OQAM}{offset quadrature amplitude modulation}
\newacronym{LMMSE}{LMMSE}{linear minimum mean square error}
\newacronym{SNR}{SNR}{signal-to-noise ratio}
\newacronym{OOBE}{OOBE}{out-of-band emission}
\newacronym{PAPR}{PAPR}{peak-to-average power ratio}
\newacronym{AFBM}{AFBM}{affine filter bank modulation}
\newacronym{FBMC}{FBMC}{filter bank multicarrier modulation}
\newacronym{PPN}{PPN}{polyphase network}
\newacronym{SIR}{SIR}{signal-to-interference ratio}
\newacronym{AF}{AF}{ambiguity function}
\newacronym{PDA}{PDA}{probabilistic data association}
\newacronym{SBL}{SBL}{sparse Bayesian learning}
\newacronym{VGA}{VGA}{vector Gaussian approximation}
\newacronym{KL}{KL}{Kullback-Leibler}
\newacronym{GAMP}{GAMP}{generalized approximate message passing}
\newacronym{EP}{EP}{expectation propagation}
\newacronym{5G}{5G}{fifth generation}
\newacronym{4G}{4G}{fourth generation}

\newcommand{\herm}[0]{^{\mathsf{H}}}

\author{
\IEEEauthorblockA{
Eya Gourar\IEEEauthorrefmark{1}\textsuperscript{\orcidlink{0009-0001-9575-6075}},
Henrique L. Senger\IEEEauthorrefmark{2}\textsuperscript{\orcidlink{0009-0004-1586-8168}},
Gustavo P. Gonçalves\IEEEauthorrefmark{2}\textsuperscript{\orcidlink{0009-0000-8260-4390}},
Kuranage R. R. Ranasinghe\IEEEauthorrefmark{3}\textsuperscript{\orcidlink{0000-0002-6834-8877}},
Hyeon Seok Rou\IEEEauthorrefmark{3}\textsuperscript{\orcidlink{0000-0003-3483-7629}},\\
Bruno S. Chang\IEEEauthorrefmark{2}\textsuperscript{\orcidlink{0000-0003-0232-7640}},
Yahia Medjahdi\IEEEauthorrefmark{1}\textsuperscript{\orcidlink{0000-0001-7030-1130}},
Giuseppe T. F. de Abreu\IEEEauthorrefmark{3}\textsuperscript{\orcidlink{0000-0002-5018-8174}},
and Didier Le Ruyet\IEEEauthorrefmark{4}\textsuperscript{\orcidlink{0000-0002-9673-2075}}
}\\
\IEEEauthorblockA{
\IEEEauthorrefmark{1} IMT Nord Europe, Institut Mines T\'el\'ecom, Center for Digital Systems, F-59653 Villeneuve d’Ascq, France\\
\IEEEauthorrefmark{2} Federal University of Technology -- Paraná, CPGEI/Electronics Department, Curitiba, Brazil\\
\IEEEauthorrefmark{3} Constructor University, School of Computer Science and Engineering, Bremen, Germany\\
\IEEEauthorrefmark{4} CEDRIC, Conservatoire National des Arts et Métiers, Paris, France\\
Email: \ eya.gourar@imt-nord-europe.fr
}
}

\begin{document}

\title{On the Robustness of AFBM Sensing to Power Amplifier Nonlinearities}

\maketitle

\begin{abstract} 
We investigate the impact of \ac{PA} nonlinearities on the sensing performance of \ac{AFBM}. 
While \ac{AFBM} offers several advantageous properties for \ac{ISAC} --including reduced \ac{OOBE}, low \ac{PAPR}, and natural robustness to \ac{DD} channel effects -- mitigating waveform distortion typically requires highly linear \acp{PA}. 
This creates a fundamental contradiction with \ac{ISAC} applications, which demand high transmit power for reliable sensing. 
Our analytical results reveal that the structure of the effective \ac{AFBM} modulation matrix dictates how distortion propagates within the \ac{AF}. 
Furthermore, simulations demonstrate that both the \ac{AF} and the overall sensing performance of \ac{AFBM} remain remarkably insensitive to such nonlinearities. 
These findings highlight the robustness of \ac{AFBM}, making it a highly viable candidate for practical \ac{ISAC} deployments constrained by hardware impairments.
\end{abstract}


\IEEEpeerreviewmaketitle

\glsresetall

\section{Introduction}

\Ac{ISAC} constitutes a paradigm shift in wireless networks, aiming to simultaneously transmit information and probe the surrounding environment using a unified hardware platform and shared spectrum. 
By coalescing communication and radar functionalities, \ac{ISAC} inherently supports advanced applications such as localization, tracking, and context-aware services \cite{liu2020joint}. 
However, this convergence imposes stringent requirements on waveform design that extend beyond conventional communication metrics, such as spectral efficiency and link reliability. 
Specifically, the transmitted waveform must exhibit favorable sensing ambiguity properties, high spectral localization, and resilience to practical hardware impairments.

\Ac{OFDM} is anticipated to remain a dominant multicarrier framework in \ac{6G} systems, largely due to its algorithmic simplicity and mature ecosystem. 
Nevertheless, \ac{OFDM} is fundamentally afflicted by a high \ac{PAPR}, a vulnerability that is particularly detrimental in \ac{ISAC} deployments. 
Elevated \ac{PAPR} pushes the \acp{PA} to operate in their nonlinear regimes which inadvertently induces severe waveform distortion. 
Extensive literature has documented the deleterious impact of \ac{PA} nonlinearities on \ac{OFDM} sensing performance (see, e.g., \cite{feng2024analysis,ismail2024robustness,akca2024integrated,gourar2025ambiguity}), demonstrating both theoretically and empirically that \ac{PA}-induced distortions elevate the ranging sidelobes of the \ac{AF}, thereby degrading radar target detection and parameter estimation. 
Consequently, there is a compelling need to investigate alternative multicarrier waveforms that preserve transceiver feasibility while ensuring robust sensing performance under hardware nonlinearities.

In this context, \ac{AFDM} has recently gained traction as a robust modulation scheme, leveraging affine (chirp-based) transformations to achieve full diversity in doubly dispersive channels \cite{rou2024orthogonal}. 
Building upon this mathematical framework, \ac{AFBM}~\cite{ranasinghe2025affinefilterbankmodulation} was introduced, harmonizing affine-domain spreading with subcarrier-wise filtering. 
This hybrid architecture inherently enhances \ac{AFDM} by drastically suppressing \ac{OOBE} and reducing \ac{PAPR}, while maintaining a manageable transceiver complexity.

Despite the inherent \ac{PAPR} advantages of \ac{AFBM}, distortions induced by nonlinear \acp{PA} remain practically inevitable, particularly in high-mobility scenarios such as vehicle-to-vehicle communications where sensitivity to \ac{RF} impairments is exacerbated. 
For sensing applications, in particular, \acp{PA} are deliberately driven near saturation to maximize the radiated power and extend radar coverage, thereby compensating for the severe attenuation of the two-way radar propagation path. 
Although implementing a large input back-off can mitigate these nonlinear effects, it incurs an unacceptable penalty on overall transmit power efficiency.

While the mathematical properties of the \ac{AFDM} \ac{AF} and the influence of chirp parameters on sidelobe suppression have been extensively characterized \cite{yin2025ambiguity,zhang2025discrete,rou2025normalized,ni2025ambiguity,bedeer2025ambiguity}, literature addressing the susceptibility of \ac{AFDM} to hardware impairments remains sparse \cite{sui2026mimo,gourar2026impact,rou2026afdm}. 
Notably, for radar sensing applications, recent analysis in \cite{gourar2026impact} demonstrated that the \ac{AFDM} \ac{AF} exhibits an inherent insensitivity to \ac{PA} nonlinearities.
This finding prompts a critical inquiry: does the filtered architecture of \ac{AFBM} preserve or enhance this resilience against nonlinear distortions?

In this paper, we answer this question by comprehensively investigating the robustness of the \ac{AFBM} waveform to \ac{PA}-induced nonlinearities and evaluating its practical viability for \ac{ISAC} deployments.
Our primary contributions are twofold. 
First, we analytically characterize the \ac{AF} of a nonlinearly amplified \ac{AFBM} signal utilizing the Bussgang decomposition, culminating in a semi-analytical expression for the average \ac{AF} power. 
Second, we assess system-level sensing performance under realistic hardware constraints.
Simulation results reveal that \ac{AFBM} is fundamentally insensitive to \ac{PA}-induced distortion, maintaining its ambiguity properties. 
Furthermore, under highly impaired hardware operation, \ac{AFBM} outperforms \ac{AFDM} in radar parameter estimation accuracy, cementing its potential as a premier waveform for future \ac{ISAC} systems.

The remainder of this paper is organized as follows: Section~\ref{section1} introduces the system model and details the construction of the \ac{AFBM} transmit signal. 
Section~\ref{section2} provides the theoretical analysis of the \ac{AF} under \ac{PA} nonlinearities. 
Finally, Section~\ref{section3} presents the numerical and simulation results, followed by our conclusions.

\section{System Model}\label{section1}


\subsection{Transmit Signal Model}

In this section we describe the \ac{AFBM} transmission chain, mainly comprising a \ac{DAFT}-spread \ac{AFDM} block followed by a polyphase network.
Let $L$ denote the number of active subcarriers in an \ac{AFBM} system with $N$ total subcarriers.
The system is organized into blocks of $K$ symbols, each with a duration of $T/2$ seconds to maintain the same rate of conventional \ac{OFDM}/\ac{AFDM} systems.
Each subcarrier is spaced by $F$ Hz. 
This results in a time-frequency grid with $L$ points in frequency, spaced by $F$ Hz, and $K$ points in time, spaced by $T/2$ seconds. 
%
%

The transmission matrix $\bar{\mathbf{G}}^{'}\in \mathbb{C}^{ON \times L}$ for a single multicarrier symbol is given by
\begin{equation}
\bar{\mathbf{G}}^{'} = \mathbf{\tilde{G}}\mathbf{Q}_{P}\mathbf{C}_f,
\label{gbartilde}
\end{equation}
where $\mathbf{\widetilde{G}}\in \mathbb{R}^{ON \times N}$ denotes the filtering matrix from a prototype filter of length $ON$, with $O$ denoting the overlap factor,   corresponding to the transmission of a single multicarrier symbol~\cite{ranasinghe2025affinefilterbankmodulation}, and 
$\mathbf{Q}_{P}\in \mathbb{C}^{N \times L}$ is comprised of an extended \ac{IDAFT} whose output of size $P$ is zero-padded in the frequency domain to $N$, obtained as
\begin{equation}
\mathbf{Q}_{P} = \mathbf{F}_{N}^{H}\mathbf{T}_{NP}\mathbf{F}_P \mathbf{\tilde{W}}\herm_{P},
\label{eq:Q_P}
\end{equation}
where $\mathbf{F}_{L}$ denotes the normalized $L$-point \gls{DFT} matrix, 
%
and
$\mathbf{T}_{NZ}\triangleq \begin{bmatrix} [\mathbf{I}_{Z/2}  \;  \mathbf{0}_{Z/2} ]^T& \mathbf{0}_{Z \times (N-Z)} &[  \mathbf{0}_{Z/2} \; \mathbf{I}_{Z/2} ]^T \end{bmatrix}^{T}$  
is an $N \times Z$ matrix with  $\mathbf{T}_{NZ}^{T}\mathbf{T}_{NZ} = \mathbf{I}_Z$. $\mathbf{0}_{Z}$ denotes a zero matrix of size $Z\times Z$, and $\mathbf{0}_{Z \times Z'}$ expresses a zero matrix of size $Z \times Z'$.
$\mathbf{W}_{L} \in \mathbb{C}^{L \times L}$ is the $L$-point \gls{DAFT} matrix, defined as
\begin{equation}
\mathbf{W}_{L} = \mathbf{\Lambda}_{c_2,L}\mathbf{F}_{L}\mathbf{\Lambda}_{c_1,L},
\end{equation}
with
\begin{equation}
\mathbf{\Lambda}_{c_i,L} = \mathrm{diag}[e^{-j2\pi c_i (0)^2}, \dots, e^{-j2\pi c_i (L-1)^2}] \in \mathbb{C}^{L \times L},
\end{equation}
denoting an $L \times L$ diagonal chirp matrix with central digital frequency $c_i$. 
The extended \ac{DAFT} $\mathbf{\tilde{W}}_{P} \in  \mathbb{C}^{L \times P}$ 
with length $P$ is defined as
\begin{equation}
\mathbf{\tilde{W}}_{P} = 
\begin{bmatrix}
\mathbf{I}_L &  \mathbf{0}_{L\times (P-L)}  
\end{bmatrix}
\mathbf{W}_P.
\label{dft_espalhada}
\end{equation}

The precoding matrix $\mathbf{C}_f\in \mathbb{C}^{L \times L}$ must be chosen to satisfy the following condition in order to restore complex orthogonality:
\begin{equation}
\mathbf{C}\herm_f\mathbf{Q}_{P}^{H}\mathbf{\tilde{G}}^{H}\mathbf{\tilde{G}}\mathbf{Q}_{P}\mathbf{C}_f \approx \mathbf{U},
\label{eq:complex_orthogonality}
\end{equation}
where $\mathbf{U} \in \mathbb{R}^{L \times L}$ is a diagonal matrix with ones in some positions of the main diagonal.
To accomplish this, let us define  $\mathbf{C}_f $ as
\begin{equation}
\mathbf{C}_f \triangleq \mathbf{W}_{L} \mathrm{diag}\{\tilde{\mathbf{b}}\}.
\label{ferf44}
\end{equation}

By substituting \eqref{ferf44} into \eqref{eq:complex_orthogonality}, the $\tilde{l}$-th element of $\tilde{\mathbf{b}}$ is obtained as
\begin{equation}
[\mathbf{\tilde{b}}]_{\tilde{l}} = 
\begin{cases} 
\sqrt{\frac{1}{[\mathbf{\tilde{c}}]_{\tilde{l}}}}, & \tilde{l} \in \left[ 0,\ldots,\tfrac{L}{4}-1 \right] \cup \left[ L-\tfrac{L}{4},\ldots,L-1 \right] \\[1ex]
0, & \text{otherwise},
\end{cases}
\end{equation}
with
\begin{equation}
\mathbf{\tilde{c}} \triangleq \mathrm{diag}\{\mathbf{W}\herm_L\mathbf{Q}_{P}^{H}\mathbf{\widetilde{G}}^H\mathbf{\widetilde{G}}\mathbf{Q}_{P}\mathbf{W}_L\}.
\end{equation}

%

The compensation stage in the precoding matrix $\tilde{\mathbf{b}}$ comprises a multiplicative factor that cancels the interference in the transmitted symbols introduced by the filter coefficients when $O \leq 1.5$. 
%
%
%
By analysing the main diagonal of $\mathbf{U}$ we will see that it has unit values only in $L/2$ positions (the first and last $L/4$ ones), and zeros elsewhere. This is consistent with other precoded \ac{FBMC} systems. 
%

$\mathbf{x} \in \mathcal{D}^{K\frac{L}{2} \times 1}$ denotes the complex transmit symbols vector for the proposed scheme. For the sake of convenience, $L = N/2$ is adopted.
Symbols in $\mathbf{x}$ are arranged in the first and last $L/4$ positions of a matrix $\mathbf{A} \in \mathbb{C}^{L \times K}$ to uphold the condition given in~\eqref{eq:complex_orthogonality}. This mapping $\bm{\Xi}$ is expressed as
\begin{equation}
\label{eq:positions}
\mathbf{a} \triangleq \mathrm{vec}(\mathbf{A}) = \bm{\Xi} \mathbf{x} \in \mathbb{C}^{LK \times 1},
\end{equation}
where $\mathrm{vec}(\cdot)$ denotes the column-wise vectorization operation and $\bm{\Xi} \in \mathbb{C}^{LK \times K\frac{L}{2}}$ is defined as
\begin{equation}
\label{eq:Xi}
\bm{\Xi} \triangleq \mathbf{I}_K \otimes \bar{\bm{\Xi}},
\end{equation}
with $\bar{\bm{\Xi}} \in \mathbb{C}^{L \times \frac{L}{2}}$ given by
\begin{equation}
\bar{\bm{\Xi}} \triangleq 
\begin{bmatrix}
\mathbf{I}_{L/4} & \mathbf{0}_{L/4} \\
\mathbf{0}_{L/2 \times L/4} & \mathbf{0}_{L/2 \times L/4} \\
\mathbf{0}_{L/4} & \mathbf{I}_{L/4}  
\end{bmatrix}.
\end{equation}

 In all, the complete \ac{AFBM} transmit signal in the \ac{TD} for all $K$ blocks $\mathbf{s}$ can be expressed in terms of the precoding matrix $\mathbf{C}_f$ in \eqref{ferf44}, the modified \ac{IDAFT} matrix $\mathbf{Q}_P$ in \eqref{eq:Q_P}, and the block Toeplitz filter matrix $\mathbf{G} \in  \mathbb{R}^{ON + (K-1)N/2 \times NK}$  obtained from  $\mathbf{\widetilde{G}}$ ~\cite{ranasinghe2025affinefilterbankmodulation}.
From (\ref{gbartilde}) and by using Kronecker product identities, the resulting signal $\mathbf{s}$ before amplification can be written as
\begin{align}
\mathbf{s} & = \mathbf{G} \mathbf{Q} \mathbf{C} \mathbf{a}  
= \mathbf{G} \big(\mathbf{I}_{K} \otimes \mathbf{Q}_{P}\big) \cdot \big(\mathbf{I}_K \otimes \mathbf{C}_f \big) \mathbf{a} \in \mathbb{C}^{M \times 1} \nonumber \\
& = \mathbf{G} \big(\mathbf{I}_{K} \otimes \mathbf{Q}_{P} \mathbf{C}_f \big) \bm{\Xi} \mathbf{x} \nonumber \\
& = \bar{\mathbf{G}} \mathbf{x}, \label{eq:afbm_mod}
\end{align}
where $M \triangleq ON + \tfrac{N}{2}(K-1)$, and the block matrix $\mathbf{Q} \in  \mathbb{C}^{NK\times LK}$ expressing the transmission of $K$ blocks is given by
\begin{equation}
\mathbf{Q} =  \mathbf{I}_{K} \otimes \mathbf{Q}_{P}.
\end{equation}
Finally, the overall modulation matrix $\bar{\mathbf{G}}$ can be expressed as $\bar{\mathbf{G}} = 
\mathbf{G}\,\mathbf{Q}\,\mathbf{C}\,\bm{\Xi}$.



Analyzing the proposed system model scheme, the proposed waveform can be interpreted as a filtered version of the DAFT-spread \ac{AFDM} scheme~\cite{tao2025affine}, where the standard sinc-chirp subcarriers are replaced with chirp-filtered subcarriers. 
Here, the considered filter is well localized both in time and in frequency and implemented with a filterbank. 
Given this well-localized filter structure, a \ac{CP} is not used in \ac{AFBM}. Before transmission, the signal is amplified; however, the nonlinearity of the transmitter’s \ac{PA} introduces distortion into the amplified signal.

\subsection{Power amplifier nonlinearities}

In this paper, we use the \ac{SSPA} model. 
Generally, a \ac{PA} model can be described by its transfer function $g(\cdot)$.

\subsubsection{Solid state power amplifier (SSPA)}
For this impairment model, also known as the \textit{Rapp} model, the $n^\text{th}$ element of the amplified signal can be written as \cite{bouhadda2014theoretical}
\begin{equation} y(n) = g(s(n)) =s(n)\left [{1 + \left ({\frac {|s(n)|}{V_{sat}}}\right) ^{2q }}\right ]^{-\frac{1}{2q }},\label{eq:rapp} \end{equation} 
where $q$ is the smoothness factor that controls the transition from linear to saturation domain determined by the input saturation voltage $V_{sat}$.

In practice, to mitigate the impacts of the nonlinear distortion, the \ac{PA} operates at an \ac{IBO} from a given saturation level. In this work, we adopt the definition
\begin{equation}
    \text{IBO} =  \frac{V_{sat}}{\sigma_s},
\end{equation}
where $\sigma_s^2=\mathbb{E}[s(n)^2]$ is the mean input signal power.

\subsubsection{Nonlinear distortion modeling}
The \ac{TD} \ac{AFBM} transmit signal is obtained following \eqref{eq:afbm_mod} from \ac{i.i.d.} zero-mean data symbols $\mathbf{x}$ drawn from a finite rotationally symmetric constellation. For sufficiently large $L$, the \ac{CLT} implies that the pre-filtered signal converges in distribution to a circular complex Gaussian random vector with covariance $\sigma_s^2 \mathbf I_M$. 
The application of the per-subcarrier filtering matrix $\mathbf{G}$ introduces correlation among the samples; however, because $\mathbf{G}$ represents linear transformations, the resulting signal remains jointly complex Gaussian \cite{gallager2008circularly}.

Hence, we approximate the \ac{PA} output using the Bussgang decomposition as \cite{bouhadda2014theoretical}
\begin{equation}
\mathbf{y} = \kappa \mathbf{s} + \mathbf{d} = \kappa \bar{\mathbf{G}} \mathbf x + \mathbf d,
\label{eq:y_afbm}
\end{equation}
where $\mathbf{d}$ denotes the nonlinear distortion with variance $\sigma_d^2 = \mathbb{E}[d(n)^2]$, uncorrelated with the input signal $\mathbf{s}$, and $\kappa$ is the complex Bussgang gain given by
\begin{equation}
\kappa = \frac{\mathbb{E}[\mathbf{s}^{\ast}\mathbf{y}]}{\mathbb{E}[|\mathbf{s}|^2]}.
\label{eq:kappa_afbm}
\end{equation}

\subsection{Received Signal}

The amplified signal vector $\mathbf{y}$ is then transmitted over a time-varying multipath channel, i.e., the doubly-dispersive channel, described concisely by the circular convolutional matrix form $\mathbf{H} \in \mathbb{C}^{M \times M}$ \cite{ranasinghe2025affinefilterbankmodulation}. 
Consequently, the received signal $\mathbf{r} \in\mathbb{C}^{M \times 1}$ is expressed by 
\begin{equation}
    \mathbf{r} = \mathbf{H} \mathbf{y} + \mathbf{n},
    \label{eq:r}
\end{equation}
where $\mathbf{n} \sim \mathcal{CN}(\mathbf{0}, \sigma_n^2 \mathbf{I}_M)$ represents additive white Gaussian noise (AWGN).

\section{Analysis of the Ambiguity Function} \label{section2}
To maximize the sensing SNR, the receiver usually processes the received signal in \eqref{eq:r} with a matched filter, where delays and Doppler frequency shifts will be observed. 
To characterize the mismatch between the waveform and the matched filter under different delay and Doppler-shift conditions, the ambiguity function (AF) is defined.

\subsection{Definition}
For a discrete-time signal $\mathbf{s}$, the AF is defined as,
\begin{equation}
\mathcal{A}(l,\nu)
=
\sum_{n} s(n) s^\ast(n-l) e^{-j2\pi \nu n},
\label{eq:AF}
\end{equation}
where $l$ denotes the delay index and $\nu$ denotes the normalized Doppler frequency. Using vector notation, the AF can be expressed in matrix form as \cite{bedeer2025ambiguity},
\begin{equation}
\mathcal{A}(l,\nu) =
\mathbf{s}^H \mathbf{D}_{\nu} \mathbf{J}_l \mathbf{s},
\label{eq:AF_matrix}
\end{equation}
where $\mathbf{J}_l$ is the delay shift matrix and $\mathbf{D}_\nu$ is the Doppler modulation matrix given by $ \mathbf{D}_\nu = \mathrm{diag}\left( 1, e^{-j2\pi\nu}, \ldots, e^{-j2\pi\nu(M-1)} \right)$.

By substituting the signal model in \eqref{eq:afbm_mod} into \eqref{eq:AF_matrix}, and defining the ambiguity matrix $\mathbf{\Phi}_{l,\nu} =\bar{\mathbf{G}}^H \mathbf{D}_\nu \mathbf{J}_l \bar{\mathbf{G}}$, the AF can be rewritten compactly as
\begin{equation}
\mathcal{A}(l,\nu)
=
\mathbf{x}^H
\mathbf{\Phi}_{l,\nu}
\mathbf{x}.
\label{eq:AF_qaudratic}
\end{equation}

For random signaling, the AF is evaluated in accordance with its average squared magnitude, i.e.,
\begin{equation}
    \bar{\mathcal{A}}(l, \nu) = 10 \operatorname{log}_{10} \mathbb{E} \big[ |\mathcal{A}(l,\nu)|^2\big], 
\end{equation}
given that the expected squared magnitude $\mathbb{E}\big[ |\mathcal{A}(l,\nu)|^2\big]$ can be calculated with \cite{bedeer2025ambiguity}
\begin{align}
\mathbb{E}\big[|\mathcal{A}(l,\nu)|^2\big]
&= \sigma_x^4\Big(|\operatorname{tr}(\mathbf{\Phi}_{l,\nu})|^2 + \|\mathbf{\Phi}_{l,\nu}\|_F^2\Big) \nonumber \\
&\quad + (\mu_4 - 2\sigma_x^4 )\| \operatorname{diag}(\mathbf{\Phi}_{l,\nu}) \|^2_2,
\end{align}
where $\sigma_x^2$ and $\mu_4$ are the second-order and fourth-order moments of the constellation symbols $\mathbf{x}$. We note that $|\operatorname{tr}(\mathbf{\Phi}_{l,\nu})|^2  = \!\sum_{i,k} \mathbf{\Phi}_{l,\nu}(i,i)\,\mathbf{\Phi}_{l,\nu}^\ast(k,k) $, $\|\mathbf{\Phi}_{l,\nu}\|_F^2 = \sum_{i,j} | \mathbf{\Phi}_{l,\nu}(i,j)|^2 $ is the Frobenius norm of the matrix $\mathbf{\Phi}_{l,\nu}$, and $\| \operatorname{diag}(\mathbf{\Phi}_{l,\nu}) \|^2_2=\sum_{n} |\mathbf{\Phi}_{l,\nu}(n,n)|^2$.

\subsection{Ambiguity function under Power amplifier nonlinearities}
Although the modulation matrix $\bar{\mathbf{G}}$ is non-unitary, \textit{i.e.}, $\bar{\mathbf{G}}\bar{\mathbf{G}}^H \neq \mathbf{I}$, its columns are approximately orthonormal, such that $\bar{\mathbf{G}}^H \bar{\mathbf{G}} \approx \mathbf{I}.$ This becomes more accurate for well-localized prototype filters and small overlap factors $O$, where inter-symbol interference is controlled.
Starting from the Bussgang decomposition in \eqref{eq:y_afbm}, we model the component of the distortion that lies in the modulation space through a distortion vector $\mathbf t$ such that $\mathbf d \approx \bar{\mathbf{G}} \mathbf t$. Equivalently, $\mathbf t$ may be interpreted as the projection of $\mathbf d$ onto the $\mathrm{DAFT}$-domain, namely $
\mathbf t \triangleq \bar{\mathbf{G}}^H \mathbf d.$
Under this approximation, the amplified signal can be written as
\begin{equation}
\mathbf y \approx \bar{\mathbf{G}}(\kappa \mathbf x + \mathbf t).
\label{eq:y_mod_domain}
\end{equation}

\noindent \textit{Remark: }For sufficiently large system dimensions, each component of $\mathbf t$ is a weighted sum of many random variables. Only if the prototype filter is well localized, \textit{i.e.,} approximately \ac{i.i.d.} distortion samples, by the \ac{CLT}, $\mathbf t$ can therefore be approximated as a zero-mean complex Gaussian random vector, with covariance matrix $\mathbf{R}_t$, and uncorrelated with $\mathbf{x}$.

Using \eqref{eq:y_mod_domain}, the \ac{AF} of the amplified signal becomes
\begin{align}
\mathcal{A}_y(l,\nu)
&= |\kappa|^2\,\mathbf{x}^H\mathbf{\Phi}_{l,\nu}\mathbf{x}
+\kappa\,\mathbf{x}^H\mathbf{\Phi}_{l,\nu}\mathbf{t}
+\kappa^{\ast}\,\mathbf{t}^H\mathbf{\Phi}_{l,\nu}\mathbf{x} \nonumber \\ 
& \quad+\mathbf{t}^H\mathbf{\Phi}_{l,\nu}\mathbf{t}. 
\label{eq:chi_y}
\end{align}

\begin{figure*}[!t]
\begin{align} 
\left| \mathcal{A}_y(l,\nu) \right|^2
&= |\kappa|^4 
\left| \mathbf{x}^H \mathbf{\Phi}_{l,\nu} \mathbf{x} \right|^2
+ |\kappa|^2
\left(
\left| \mathbf{x}^H \mathbf{\Phi}_{l,\nu} \mathbf{t} \right|^2
+
\left| \mathbf{t}^H \mathbf{\Phi}_{l,\nu} \mathbf{x} \right|^2
\right)
+ \left| \mathbf{t}^H \mathbf{\Phi}_{l,\nu} \mathbf{t} \right|^2\nonumber\\
&\quad
+ 2\,\Re \Big\{
|\kappa|^2 \kappa^\ast
\left( \mathbf{x}^H \mathbf{\Phi}_{l,\nu} \mathbf{x} \right)
\left( \mathbf{t}^H \mathbf{\Phi}_{l,\nu}^H \mathbf{x} \right)
+ |\kappa|^2 \kappa
\left( \mathbf{x}^H \mathbf{\Phi}_{l,\nu} \mathbf{x} \right)
\left( \mathbf{x}^H \mathbf{\Phi}_{l,\nu}^H \mathbf{t} \right)
+ |\kappa|^2
\left( \mathbf{x}^H \mathbf{\Phi}_{l,\nu} \mathbf{x} \right)
\left( \mathbf{t}^H \mathbf{\Phi}_{l,\nu}^H \mathbf{t} \right)\nonumber\\
&\qquad \qquad 
+ \kappa^2
\left( \mathbf{x}^H \mathbf{\Phi}_{l,\nu} \mathbf{t} \right)
\left( \mathbf{x}^H \mathbf{\Phi}_{l,\nu}^H \mathbf{t} \right)
+ \kappa
\left( \mathbf{x}^H \mathbf{\Phi}_{l,\nu} \mathbf{t} \right)
\left( \mathbf{t}^H \mathbf{\Phi}_{l,\nu}^H \mathbf{t} \right)
+ \kappa^\ast
\left( \mathbf{t}^H \mathbf{\Phi}_{l,\nu} \mathbf{x} \right)
\left( \mathbf{t}^H \mathbf{\Phi}_{l,\nu}^H \mathbf{t} \right)
\Big\}.\label{eq:|A(l,nu)|^2}
\end{align}
\end{figure*}
The squared magnitude of~\eqref{eq:chi_y} is written in~\eqref{eq:|A(l,nu)|^2}. It can be seen that the nonlinearities introduce distortion terms that may or may not add to the sum due to the non-fixed sign of the terms residing in the real-part operator $\Re \{\cdot\}$.
\begin{figure*}[!t]
\begin{align}
\mathbb E\!\left[|\mathcal A_y(l,\nu)|^2\right]
&\approx |\kappa|^4
\Big[
\sigma_x^4\Big(
|\operatorname{tr}(\mathbf\Phi_{l,\nu})|^2
+
\|\mathbf\Phi_{l,\nu}\|_F^2
\Big) + 
(\mu_4-2\sigma_x^4)\|
\operatorname{diag}(\mathbf\Phi_{l,\nu})
\|_2^2
\Big] +
|\kappa|^2\sigma_x^2
\operatorname{tr}\!\left(
\mathbf\Phi_{l,\nu}\mathbf R_t\mathbf\Phi_{l,\nu}^H
\right) \nonumber \\
&\quad+ 
|\kappa|^2\sigma_x^2
\operatorname{tr}\!\left(
\mathbf\Phi_{l,\nu}^H\mathbf R_t\mathbf\Phi_{l,\nu}
\right) +
\left|
\operatorname{tr}\!\left(
\mathbf\Phi_{l,\nu}\mathbf R_t
\right)
\right|^2
+
\operatorname{tr}\!\left(
\mathbf\Phi_{l,\nu}\mathbf R_t\mathbf\Phi_{l,\nu}^H\mathbf R_t
\right) +
2|\kappa|^2\sigma_x^2
\Re\!\left\{
\operatorname{tr}(\mathbf\Phi_{l,\nu})
\operatorname{tr}\!\left(
\mathbf\Phi_{l,\nu}\mathbf R_t
\right)^\ast
\right\}. 
\label{eq:E[|A(l,nu)|^2]}
\end{align}
\noindent\rule{\textwidth}{0.01pt}
\end{figure*}
The expected squared magnitude of the \ac{AF} of the amplified signal is approximated using the Bussgang decomposition in \eqref{eq:E[|A(l,nu)|^2]} (see derivation in appendix \ref{proof1}). This formulation reveals the role of the structure of the ambiguity matrix $\mathbf{\Phi}_{l,\nu}$ and the distortion statistics (through $\mathbf{R}_t$) in shaping the sidelobes under nonlinear conditions. This motivates an investigation of how nonlinearity affects the \ac{AF} and, consequently, practical sensing performance, as well as an analysis of the prototype filter’s sensitivity to such distortions.

\section{Numerical results}\label{section3}
In this section, we present simulation results of the sensing performance of the \ac{AFBM} scheme, alongside a comparison with the classical AFDM and OFDM schemes.
For these simulations, the total number of subcarriers $L$ is 128, the chirp size $P$ is 128, and the filter bank $\mathrm{DFT}$ size $N$ is 256. Each transmission consists of $ K=8$ 4-\ac{QAM} symbols. The chirp frequencies of each \ac{IDAFT}/\ac{DAFT} are appropriately selected to ensure the orthogonality condition and maintain a low \ac{PAPR} \cite{ranasinghe2025affinefilterbankmodulation,rou2024orthogonal}.
The nonlinear \ac{PA} is modeled using the \textit{Rapp} AM/AM function in \eqref{eq:rapp}, and the distortion level is controlled through the \ac{IBO}.
\subsection{Ambiguity function}
\begin{figure}[!t]
  \centering
    \subfloat[]{%
  \includegraphics[width=1\linewidth]{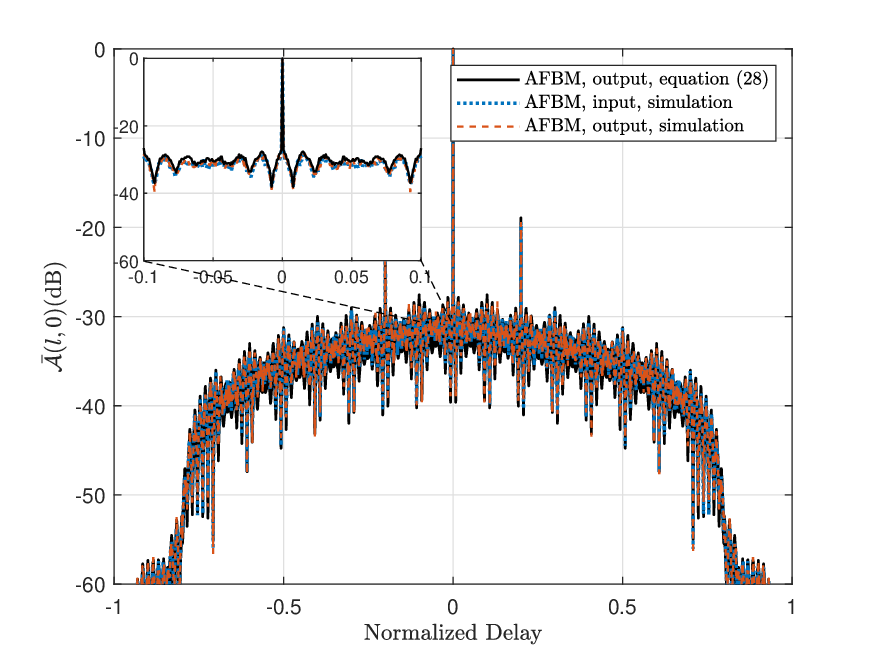}
    }\\
    \subfloat[]{%
  \includegraphics[width=1\linewidth]{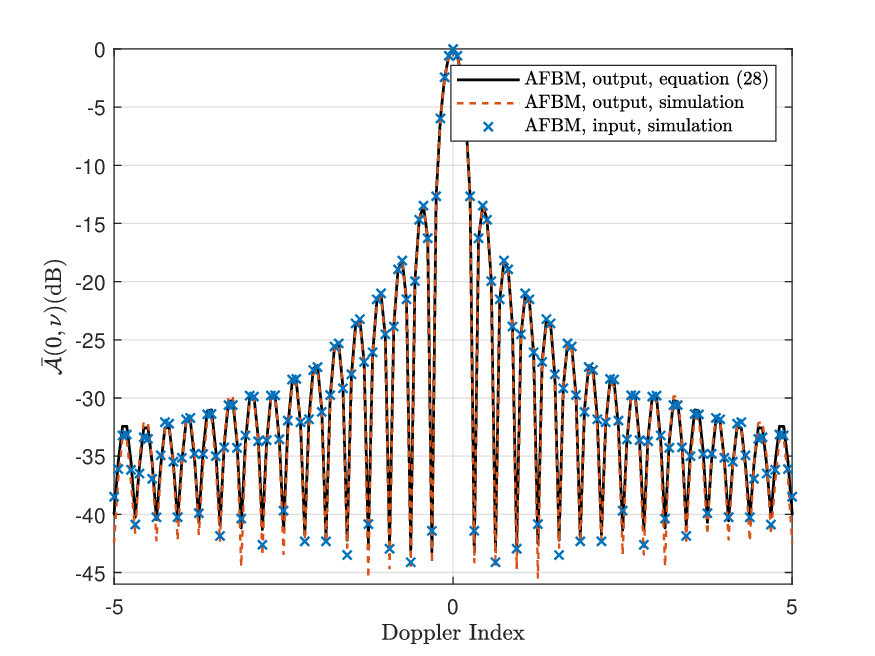}
    }
    \caption{\centering{Comparison of the (a) zero-Doppler and (b) zero-delay cuts of the AFBM, before and after nonlinear amplification, with the Phydyas filter, $O=2$ and IBO = 1~dB.}} 
    \label{fig:sim&theo}
\end{figure}

We evaluate the AFs using their normalized average performance. We first illustrate in Fig.~\ref{fig:sim&theo} the zero-Doppler cut $\mathcal{A}(l,0)$ and zero-delay cut $\mathcal{A}(0,\nu)$ of the \ac{AFBM} before and after the nonlinear amplification, using the Phydyas filter with $O=2$. We note that the overlap factor is kept low to improve the achievable \ac{SIR} \cite{ranasinghe2025affinefilterbankmodulation}. 
The zero-Doppler sidelobes appear insensitive to \ac{PA} nonlinearities, as they remain essentially unchanged, which is likely due to the fact that the original sidelobe levels are already relatively high, such that the additional terms introduced by the nonlinearities, as shown in \eqref{eq:E[|A(l,nu)|^2]}, remain negligible in comparison. Similarly, in the zero-delay (Doppler) cut, the overall sidelobe level is almost invariant. However, a decrease in the recurrent sidelobe sinks/depressions is observed. Since the considered \ac{IBO} is $1$ dB, operating at higher \ac{IBO} values, \textit{i.e.}, under less severe nonlinearities, leads to identical observations. 
We also show that the theoretical values following the approximation in \eqref{eq:E[|A(l,nu)|^2]} closely match the empirical results, with occasional mismatches in the zero-delay cut of a maximum of 3 dB. This mismatch is mainly in the sidelobe depressions, where the values are below -40~dB and therefore highly sensitive to even small variations in the distortion energy. This stems from the projection-based modeling of nonlinear distortion in Section~\ref{section2}, which omits small components that lie outside the modulation subspace. Consequently, the analytical model might slightly overestimate the AF energy. Additional discrepancies may also arise from the simplifying assumptions introduced to ensure the analytical tractability in deriving \eqref{eq:E[|A(l,nu)|^2]}.  

\begin{figure}[!t]
  \centering
    \subfloat[]{%
  \includegraphics[width=1\linewidth]{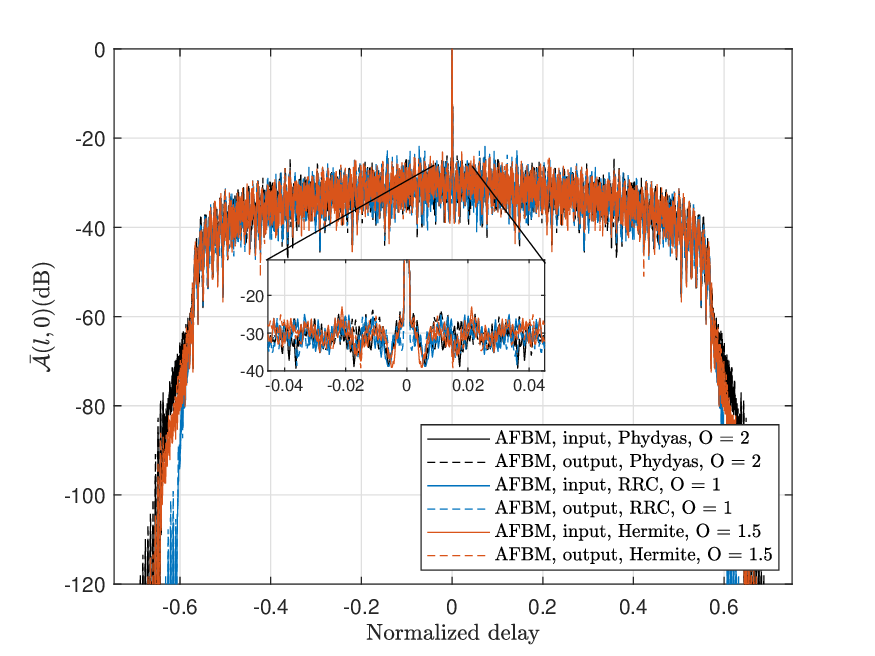}
    }\\
\subfloat[]{%
  \includegraphics[width=1\linewidth]{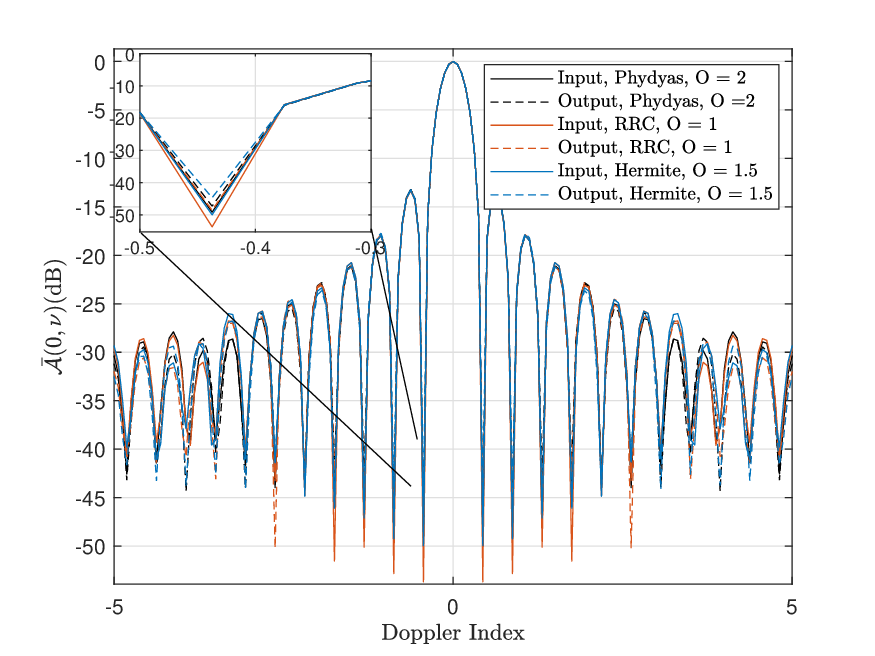}
    }
    \caption{\centering{Comparison of the (a) zero-Doppler and (b) zero-delay cuts of the AFBM, before and after nonlinear amplification, with different prototype filters. IBO = 1 dB.}} 
    \label{fig:AF_vs_Filters}
\end{figure}

Next, we present in Fig.~\ref{fig:AF_vs_Filters} the zero-Doppler and zero-delay cuts of the input and output \ac{AFBM} signals implementing different prototype filters. Given that the latter are generated with different overlap factors $O$, their effective time support may differ, thereby affecting the ranging sidelobe level \cite{liu2025cp}. Therefore, to ensure a fair comparison, filters are truncated to the same support, ideally $1.5 N$ to avoid overly truncating long filters. The results show that, regardless of the filter choice, the presence of the nonlinearities leads to similar behavior; the ranging sidelobe levels remain largely unaffected, while the Doppler sidelobes are mainly reduced at relatively high Doppler shifts.

\begin{figure}[!t]
  \centering
    \subfloat[]{%
  \includegraphics[width=1\linewidth]{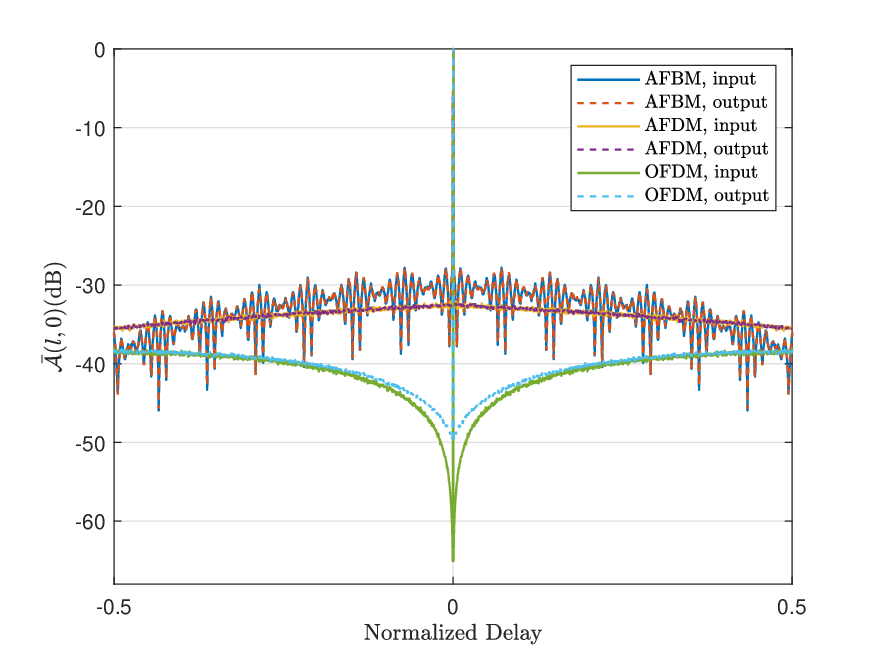}
    }\\
\subfloat[]{%
  \includegraphics[width=1\linewidth]{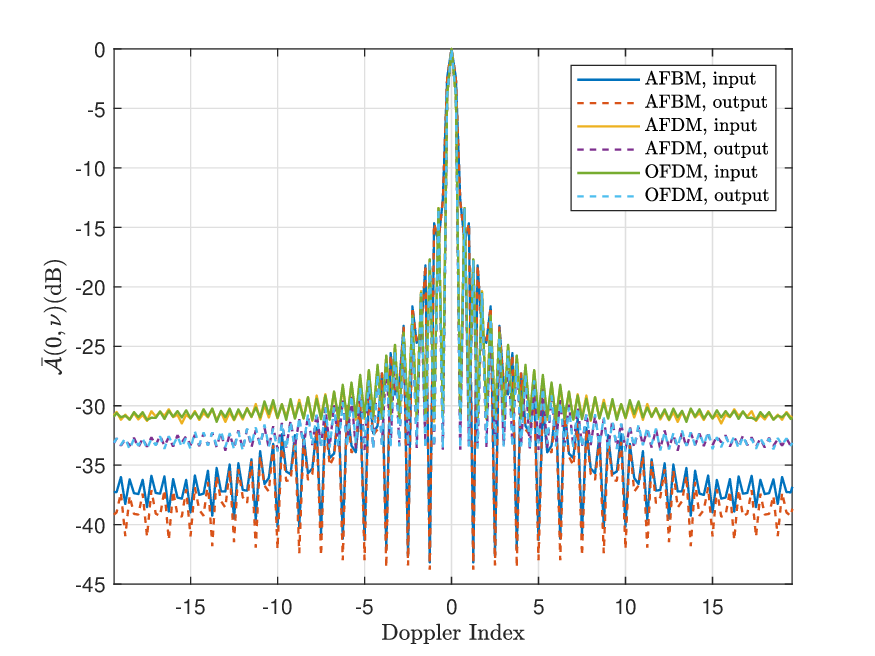}
    }
    \caption{\centering{Comparison of the (a) zero-Doppler and (b) zero-delay cuts of the AFBM with the Phydyas filter, AFDM, and OFDM, before and after nonlinear amplification.}} 
    \label{fig:AF_waveforms}
\end{figure}

In Fig.~\ref{fig:AF_waveforms}, we compare the zero-Doppler and zero-delay cuts of \ac{AFBM} with those of other candidate waveforms, namely, \ac{AFDM}, and \ac{OFDM}, both at the input and output of the nonlinear \ac{PA}. The modulation parameters of \ac{AFDM} and OFDM are selected such that the generated sequences have the same effective length $M$ as the \ac{AFBM} waveform. The \ac{AFDM} chirp parameters are set to $c_{1,\bar M} = 1/(2\bar M)$ and $c_{2,\bar M} = 1/(2 \bar M^2)$ for better localization of the \ac{AF} where $\bar M = \frac{KL}{2}$ \cite{ranasinghe2025affinefilterbankmodulation}. Meanwhile, the usage of 4-\ac{QAM}/\ac{QPSK} leads to optimal ranging sidelobes of the \ac{OFDM} \cite{liu2025cp}, and compatibility across the waveforms.
As shown in Fig.~\ref{fig:AF_waveforms}(a), the delay sidelobe levels of \ac{AFBM} are comparable to those of \ac{AFDM}, and both waveforms exhibit similar insensitivity to nonlinear distortion in the ranging (delay) domain. In contrast, OFDM experiences sidelobe regrowth mainly near the mainlobe \textit{i.e,} smaller delays, but still achieves globally lower delay sidelobes. On the other hand, the zero-delay cuts in Fig.~\ref{fig:AF_waveforms}(b) reveal that \ac{AFBM} achieves lower Doppler sidelobes compared to both \ac{AFDM} and \ac{OFDM}, whose performances are similar. The impact of nonlinearities on the latter waveforms is reflected in a prominent decrease of the sidelobes, whereas \ac{AFBM} does not exhibit the same level of decrease at the same Doppler indices and shows lesser responsiveness to such distortion.

\subsection{Sensing performance}
\begin{figure}[!t]
  \centering
  \includegraphics[width=1\linewidth]{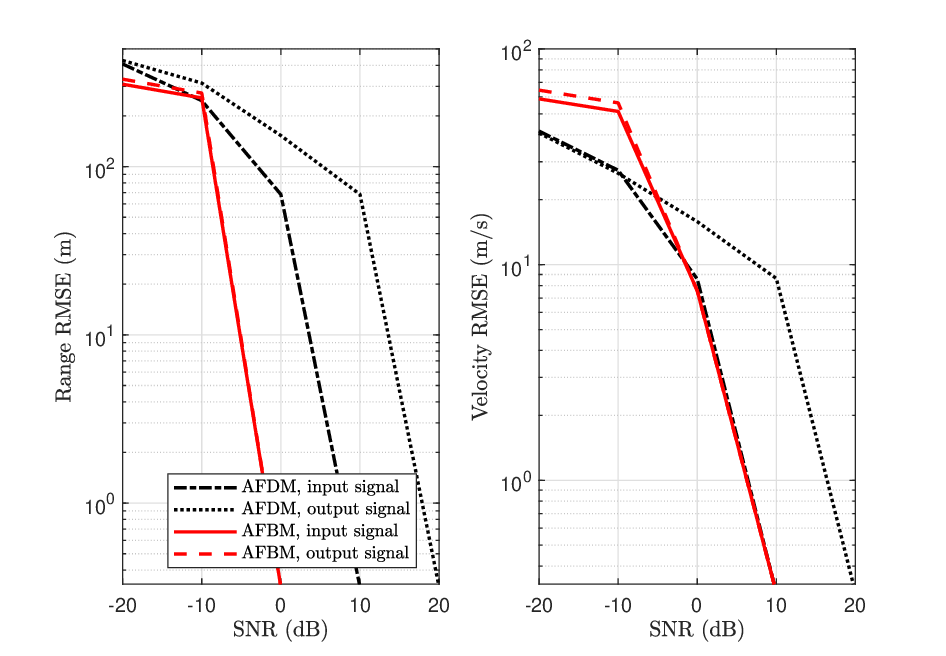}
    \caption{\centering{Radar parameter estimation performance of the candidate waveforms, in terms of RMSE for target range and velocity in both linear and nonlinear cases with IBO = 2~dB.}}
    \label{fig:RMSE}
\end{figure}

To assess the system-level sensing performance, we implement the system model described in Section~\ref{section1} over a doubly dispersive channel with three resolvable paths, each characterized by its normalized delay and digital Doppler shift. The carrier frequency $f_c $ is fixed to 4~GHz. To estimate the delay and Doppler shift alongside each path, the PDA-based approach proposed in \cite{ranasinghe2025affinefilterbankmodulation} is used. Fig.~\ref{fig:RMSE} presents the \ac{RMSE}  performance of the estimator when applied to the \ac{AFBM} and \ac{AFDM} waveforms. We note that the resolution limits are consistent with those reported in \cite{ranasinghe2025affinefilterbankmodulation}, since the same system parameters are used and the resolution is primarily determined by these parameters. While the \ac{AF} of \ac{AFBM} and \ac{AFDM} remain largely unchanged, particularly the ranging sidelobes, under \ac{PA} nonlinearities, the \ac{AFBM} waveform appears more robust across the \ac{SNR} range. This indicates that the observed performance gap is not solely explained by ambiguity sidelobes, since the PDA-based estimator is also sensitive to how nonlinear distortion projects onto the sensing dictionary.

\section{Conclusion}
In this paper, we investigated the impact of \ac{PA} nonlinearities on the \ac{AF} and sensing performance of the \ac{AFBM} waveform. Leveraging a Bussgang-based analytical framework, we characterized how nonlinear distortion propagates through the AFBM modulation structure and affects the resulting \ac{AF}. 
Our analysis, supported by numerical simulations, demonstrated that \ac{AFBM} exhibits a remarkable insensitivity to \ac{PA}-induced nonlinearities. In particular, both the \ac{AF} and the sensing metrics, such as \ac{RMSE}, remain largely unaffected even under severe hardware impairments. 
Furthermore, comparative results showed that \ac{AFBM} can outperform \ac{AFDM} in highly nonlinear regimes, highlighting the benefits of its filtered affine-domain structure.
Future work may extend this analysis to other \ac{RF} impairments on \ac{AFBM}-based sensing systems.

\appendices

\section{Approximation of the expectation of \eqref{eq:|A(l,nu)|^2}}\label{proof1}
We define $a = |\kappa|^2 \mathbf x^H\mathbf\Phi\mathbf x$, $b = \kappa \mathbf x^H\mathbf\Phi\mathbf t$,  $c = \kappa^\ast \mathbf t^H\mathbf\Phi\mathbf x$, and $d = \mathbf t^H\mathbf\Phi\mathbf t$. Hence \eqref{eq:|A(l,nu)|^2} becomes
\begin{align}
|\mathcal A_y(l,\nu)|^2 &= |a|^2+|b|^2+|c|^2+|d|^2 \nonumber \\ &\quad+2\Re\{ab^\ast+ac^\ast+ad^\ast+bc^\ast+bd^\ast+cd^\ast\}.\label{eq:E_Ay_reduced}
\end{align}
Since the input $\mathbf s$ is circular and the considered \textit{Rapp model} is phase preserving, the distortion $\mathbf d$ is also modeled as circular. Therefore, its projection $\mathbf t=\bar{\mathbf G}^H\mathbf d$ is also proper (circular) \cite{neeser2002proper}. Each element of $\mathbf{t}$ depends on a large number of data symbols through spreading and filtering, making the dependence between $\mathbf{t}$ and any individual data symbol weak. Therefore, we assume that higher-order moments involving $\mathbf{x}$ and $\mathbf{t}$ are approximated by factorizing expectations over $\mathbf{x}$ and $\mathbf{t}$\footnote{The validity of this approximation is supported by empirical results, as shown in Section~\ref{section3}.}. Since $\mathbf{t}$ is zero-mean, and $\mathbf{x}$ and $\mathbf{t}$ are uncorrelated and proper, the resulting terms contain factors such as $\mathbb E[\mathbf t]$ and third-order moments of $\mathbf{t}$ that vanish. Hence,
\begin{equation}
\mathbb E[ab^\ast] \approx \mathbb E[ac^\ast] \approx \mathbb E[bd^\ast] \approx \mathbb E[cd^\ast] \approx0.
\end{equation}
Using $\mathbb E[\mathbf x\mathbf x^H]=\sigma_x^2\mathbf I_{K\frac{L}{2}}$, the signal-distortion terms become
\begin{align}
\mathbb E[|b|^2]
&= |\kappa|^2 \mathbb E\!\left[
\mathbf x^H\mathbf\Phi\mathbf t\mathbf t^H\mathbf\Phi^H\mathbf x
\right] \approx |\kappa|^2\sigma_x^2
\operatorname{tr}\!\left(
\mathbf\Phi\mathbf R_t\mathbf\Phi^H
\right),\\
\mathbb E[|c|^2]
&=|\kappa|^2 \mathbb E\!\left[
\mathbf t^H\mathbf\Phi\mathbf x\mathbf x^H\mathbf\Phi^H\mathbf t
\right] \approx
|\kappa|^2\sigma_x^2
\operatorname{tr}\!\left(
\mathbf\Phi^H\mathbf R_t\mathbf\Phi
\right).
\end{align}
Moreover, under the above approximation, the mixed fourth-order term is omitted, i.e., $\mathbb E[bc^\ast]=0$.
Furthermore, the pure distortion term can be approximated as
\begin{equation}
\mathbb E[|d|^2]
\approx \left|
\operatorname{tr}\!\left(
\mathbf\Phi\mathbf R_t
\right)
\right|^2 +
\operatorname{tr}\!\left(
\mathbf\Phi\mathbf R_t\mathbf\Phi^H\mathbf R_t
\right).
\end{equation}
Finally,
\begin{align}
\mathbb E[ad^\ast]
&\approx 
|\kappa|^2
\mathbb E[\mathbf x^H\mathbf\Phi\mathbf x]
\,
\mathbb E[(\mathbf t^H\mathbf\Phi\mathbf t)^\ast]
\nonumber\\
&\approx 
|\kappa|^2
\left(
\sigma_x^2\operatorname{tr}(\mathbf\Phi)
\right)
\left(
\operatorname{tr}(\mathbf\Phi\mathbf R_t)
\right)^\ast.
\end{align}
Substituting the above expressions into \eqref{eq:E_Ay_reduced} yields \eqref{eq:E[|A(l,nu)|^2]}. \hfill\IEEEQED

\selectlanguage{english}
\bibliographystyle{IEEEtran}
\bibliography{biblio.bib}

@misc{gourar2025ambiguity,
      title={On the Ambiguity Function of OFDM-based ISAC Signals Under Non-Ideal Power Amplifiers}, 
      author={Eya Gourar and Yahia Medjahdi and Laurent Clavier and Abdul Karim Gizzini and Patrick Sondi},
      year={2025},
      eprint={2512.09803},
      archivePrefix={arXiv},
      primaryClass={eess.SP},
      url={https://arxiv.org/abs/2512.09803}, 
}

@article{bedeer2025ambiguity,
  title={Ambiguity function analysis of affine frequency division multiplexing for integrated sensing and communication},
  author={Bedeer, Ebrahim},
  journal={arXiv preprint arXiv:2504.02582},
  year={2025}
}

@article{liu2025cp,
  title={CP-OFDM achieves the lowest average ranging sidelobe under QAM/PSK constellations},
  author={Liu, Fan and Zhang, Ying and Xiong, Yifeng and Li, Shuangyang and Yuan, Weijie and Gao, Feifei and Jin, Shi and Caire, Giuseppe},
  journal={IEEE Transactions on Information Theory},
  year={2025},
  publisher={IEEE}
}

@article{yin2025ambiguity,
  title={Ambiguity function analysis of AFDM signals for integrated sensing and communications},
  author={Yin, Haoran and Tang, Yanqun and Ni, Yuanhan and Wang, Zulin and Chen, Gaojie and Xiong, Jun and Yang, Kai and Kountouris, Marios and Guan, Yong Liang and Zeng, Yong},
  journal={IEEE Journal on Selected Areas in Communications},
  year={2025},
  publisher={IEEE}
}

@article{zhang2025discrete,
  title={On Discrete Ambiguity Functions of Random Communication Waveforms},
  author={Zhang, Ying and Liu, Fan and Xiong, Yifeng and Yuan, Weijie and Li, Shuangyang and Zheng, Le and Han, Tony Xiao and Masouros, Christos and Jin, Shi},
  journal={arXiv preprint arXiv:2512.08352},
  year={2025}
}

@article{rou2025normalized,
  title={Normalized Ambiguity Function Characteristics of OFDM, OTFS, AFDM, and CP-AFDM for ISAC},
  author={Rou, Hyeon Seok and de Abreu, Giuseppe Thadeu Freitas},
  journal={arXiv preprint arXiv:2510.11216},
  year={2025}
}

@article{ni2025ambiguity,
  title={Ambiguity function analysis of AFDM under pulse-shaped random ISAC signaling},
  author={Ni, Yuanhan and Liu, Fan and Yin, Haoran and Tang, Yanqun and Wang, Zulin},
  journal={arXiv preprint arXiv:2511.04200},
  year={2025}
}

@inproceedings{feng2024analysis,
  title={{Analysis of Non-linear Power Amplifier Effect and Digital Predistortion on OFDM Radar}},
  author={Feng, Ruoyu and Bauduin, Marc and Bourdoux, Andr{\'e}},
  booktitle={2024 21st European Radar Conference (EuRAD)},
  pages={248--251},
  year={2024},
  organization={IEEE}
}

@inproceedings{ismail2024robustness,
  title={{Robustness of ISAC Waveforms to Power Amplifier Distortion}},
  author={Ismail, Abdur Rahman Mohamed and Guenach, Mamoun and Sakhnini, Adham and Bourdouk, Andr{\'e} and Steendam, Heidi},
  booktitle={2024 IEEE 4th International Symposium on Joint Communications \& Sensing (JC\&S)},
  pages={1--6},
  year={2024},
  organization={IEEE}
}

@inproceedings{akca2024integrated,
  title={{Integrated Sensing and Communication with Power Amplifier Impairment}},
  author={Akca, Huseyin and Memi{\c{s}}o{\u{g}}lu, Ebubekir and {\c{C}}{\i}rpan, Hakan Ali and Arslan, Huseyin},
  booktitle={2024 6th International Conference on Communications, Signal Processing, and their Applications (ICCSPA)},
  pages={1--6},
  year={2024},
  organization={IEEE}
}

@article{sui2026mimo,
  title={MIMO-AFDM Outperforms MIMO-OFDM in the Face of Hardware Impairments},
  author={Sui, Zeping and Liu, Zilong and Musavian, Leila and Guan, Yong Liang and Yang, Lie-Liang and Hanzo, Lajos},
  journal={arXiv preprint arXiv:2601.00502},
  year={2026}
}

@article{rou2026afdm,
  title={AFDM: Evolving OFDM Towards 6G+},
  author={Rou, Hyeon Seok and Savaux, Vincent and Sui, Zeping and de Abreu, Giuseppe Thadeu Freitas and Liu, Zilong},
  journal={arXiv preprint arXiv:2602.08163},
  year={2026}
}

@article{gourar2026impact,
  title={Impact of PA Nonlinearities on AFDM Sensing: A Matched Filtering Perspective},
  author={Gourar, Eya and Medjahdi, Yahia and Clavier, Laurent and Gizzini, Abdul Karim and Sondi, Patrick},
  year={2026}
}

@ARTICLE{ranasinghe2025affinefilterbankmodulation,
  author={Ranasinghe, Kuranage Roche Rayan and Senger, Henrique L. and Gonçalves, Gustavo P. and Rou, Hyeon Seok and Chang, Bruno S. and Abreu, Giuseppe Thadeu Freitas de and Le Ruyet, Didier},
  journal={IEEE Transactions on Wireless Communications}, 
  title={Affine Filter Bank Modulation (AFBM): A Novel 6G ISAC Waveform With Low PAPR and OOBE}, 
  year={2026},
  volume={25},
  number={},
  pages={12754-12769},
  doi={10.1109/TWC.2026.3666362}}

@article{neeser2002proper,
  title={Proper complex random processes with applications to information theory},
  author={Neeser, Fredy D and Massey, James L},
  journal={IEEE transactions on information theory},
  volume={39},
  number={4},
  pages={1293--1302},
  year={1993},
  publisher={IEEE}
}

@article{rou2024orthogonal,
  title={From orthogonal time--frequency space to affine frequency-division multiplexing: A comparative study of next-generation waveforms for integrated sensing and communications in doubly dispersive channels},
  author={Rou, Hyeon Seok and De Abreu, Giuseppe Thadeu Freitas and Choi, Junil and Gonz{\'a}lez, David and Kountouris, Marios and Guan, Yong Liang and Gonsa, Osvaldo},
  journal={IEEE Signal Processing Magazine},
  volume={41},
  number={5},
  pages={71--86},
  year={2024},
  publisher={IEEE}
}

@article{tao2025affine,
author={Tao, Yiwei and Wen, Miaowen and Ge, Yao and Mao, Tianqi and Tang, Yanqun and Doosti-Aref, Abed},
  journal={IEEE Transactions on Wireless Communications}, 
  title={Affine Frequency Division Multiple Access Based on DAFT Spreading for Next-Generation Wireless Networks}, 
  year={2026},
  volume={25},
  number={},
  pages={4626-4641},
  doi={10.1109/TWC.2025.3612880}}

@article{bouhadda2014theoretical,
  title={{Theoretical analysis of BER performance of nonlinearly amplified FBMC/OQAM and OFDM signals}},
  author={Bouhadda, Hanen and Shaiek, Hmaied and Roviras, Daniel and Zayani, Rafik and Medjahdi, Yahia and Bouallegue, Ridha},
  journal={EURASIP Journal on Advances in Signal Processing},
  volume={2014},
  number={1},
  pages={60},
  year={2014},
  publisher={Springer}
}

@article{gallager2008circularly,
  title={Circularly-symmetric Gaussian random vectors},
  author={Gallager, Robert G},
  journal={preprint},
  volume={1},
  year={2008}
}

@article{liu2020joint,
  title={Joint radar and communication design: Applications, state-of-the-art, and the road ahead},
  author={Liu, Fan and Masouros, Christos and Petropulu, Athina P and Griffiths, Hugh and Hanzo, Lajos},
  journal={IEEE Transactions on Communications},
  volume={68},
  number={6},
  pages={3834--3862},
  year={2020},
  publisher={IEEE}
}

\end{document}